\documentclass[%
 reprint,
 amsmath,amssymb,
 aps,
]{revtex4-1}

\usepackage{graphicx}
\usepackage{dcolumn}
\usepackage{bm}
\usepackage{subfigure}

\begin{document}


\title{The freezing R\`{e}nyi quantum discord}

\author{Xiao-Yu Li $^{1}$}
 \altaffiliation{xiaoyu33521@163.com}
\author{Qin-Sheng Zhu $^{2}$}%
 \email{zhuqinsheng@gmail.com}
\author{Ming-Zheng Zhu $^{2}$}%
\author{Hao Wu $^{1}$}%
\author{Shao-Yi Wu $^{1}$}%
\author{Ming-Chuan Zhu $^{1}$}%
\affiliation{%
  $^{1}$ School of information and software engineering, University of Electronic Science and Technology of China, Chengdu, 610054, P.R.China
}%
\affiliation{%
  $^{2}$ School of Physical, University of Electronic Science and Technology of China, Chengdu, 610054, P.R.China
}%

\date{\today}

\begin{abstract}
 As a universal quantum character of quantum correlation, the freezing phenomenon is researched by geometry and quantum discord methods, respectively. In this paper, the properties of R\`{e}nyi discord is studied for two independent Dimer System coupled to two correlated Fermi-spin environments under the non-Markovian condition. We further demonstrate that the freezing behaviors still exist for R\`{e}nyi discord and study the effects of different parameters on this behaviors.
\begin{description}
\item[PACS numbers]
03.65.Yz, 03.65.Ud, 02.30.Yy, 03.67.-a.
\item[Key words]
R\`{e}nyi discord; the freezing phenomenon; quantum correlation.
\end{description}
\end{abstract}

\pacs{Valid PACS appear here}
\maketitle


\section{\label{sec:level1}Introduction}

As an important part of the quantum theory,  the quantum correlation has aroused extensive attention in
lots of physical fields, such as quantum information[1-3], condensed matter physics [4-5] and
gravitation wave [6] due to some unimaginable properties in a composite
quantum system which can not be reproduced by a classical system. In the past twenty years,
entanglement was considered as the quantum correlation and gradually understood. But,
the quantum discord concept has been put forward by Ollivier and Zurek [7] and Henderson and
Vedral [8] with the deep understanding of quantum correlation. It was clearly demonstrated that
entanglement represents only a portion of the quantum correlations and can entirely cover the
 latter only for a global pure state [9]. Later, many efforts have been devoted to quantify
 quantum correlation from the view of geometry [9-15] and entropy [7-8,16-19].

Since the systematic correlation contains two parts: the classical correlations and quantum
correlation, Maziero et al. [20] found the frozen behavior of the classical correlations for
phase-flip, bit-flip, and bit-phase flip channels. As for the possible similar behaviors for the quantum correlation,
Mazzola, Piilo, and Maniscalco [21] displayed the similar behavior of the quantum
correlations under the nondissipative-independent-Markovian reservoirs for special choices of
the initial state. In the same year, Lang and Caves [22] provided a complete geometry picture
of the freezing discord phenomenon for Bell-diagonal states. Later, some effort has been devoted
to discuss the condition for the frozen-discord with some Non-Markovian processes and inial
states[9,14-15,23] ( Bell-diagonal states, X states and SCI atates). In  conclusion, the freezing discord
shows a robust feature of a family of two-qubit models subject to nondissipative decoherence.
Although different measures of discords lead to some different conditions for the freezing
phenomenon, this phenomenon of quantum correlation reflects a deeper physical interpretation,
such as some relationship with quantum phase transition [24].

Recently, the R\`{e}nyi entropy
\begin{eqnarray}
S_{\alpha}(\rho)=\frac{1}{1-\alpha}\log Tr[\rho ^{\alpha}]
\end{eqnarray}
arouses much attention because it is easier to implement in the experiment than the von Neumann entropy
which needs the tool of tomography.  Here the parameter $\alpha\in(0,1)\cup(1,\infty)$ and the logarithm
is in base 2. Notably, the R$\acute{e}$nyi entropy will reduce to the von Neumann entropy when $\alpha\rightarrow 1$.
As an natural extension of quantum discord, the R\`{e}nyi entropy discord (RED)[25] is also put forward.
Therefore, it is valuable to study the properties of RED and the
condition for the freezing phenomenon of RED in quantum
information field.

\section{\label{sec:level2}The quantum correlation of Dimer system}
\subsection{\label{sec:level2}The definition of R\`{e}nyi discord}
At first, Ollivier and Zurek [7] gave the concept of quantum discord (QD)
\begin{eqnarray}
D(\rho_{AB})=\mathop{min}_{\Pi_{k}^{A}}\sum_{k}p_{k}S(\rho_{k}^{B})+S(\rho_{B})-S(\rho_{AB})
\end{eqnarray}
to quantify the quantum correlation, where the von Neumann entropy $S(X)=-tr(\rho_{X}\log_{2} \rho_{X})$ is for the density operator $\rho_{X}$ of system $X$, $\rho_{A(B)}=Tr_{B(A)}(\rho_{AB})$ is the reduced density matrix by tracing out the degree of the system $B(A)$, $p_{k}=Tr((\Pi_{k}^{A})^{\dag}\rho_{AB}\Pi_{k}^{A})$ and $\rho_{k}^{B}=Tr_{A}((\Pi_{k}^{A})^{\dag}\rho_{AB}(\Pi_{k}^{A})/p_{k}$.

Later, an equivalent description is introduced in ref [25-26]. The main idea of this equivalent description is to apply an isometry extension of the measurement map $U_{A\rightarrow EX}$ from $A$ to a composite system $EX$. This method reveals that any channel from $A$ to $A'$ can be used to describe the composite system $EX$ when we discard the freedom of $E$. Finally, the quantum discord is rewritten as:
\begin{eqnarray}
D(\rho_{AB})=\mathop{inf}_{\Pi_{k}^{A}}I(E;B\mid X)_{\tau_{XEB}}
\end{eqnarray}
where the optimization is with respect to all possible POVMs $\Pi_{k}^{A}$ of system A with the classical output X. E is an environment for the measurement map and
\begin{eqnarray}
\tau_{XEB}&=&U_{A\rightarrow EX}\rho_{AB}U_{A\rightarrow EX}^{\dagger}\nonumber\\
U_{A\rightarrow EX}\mid\psi_{A}\rangle&=&\sum_{k}|k\rangle_{X}\otimes (\sqrt{\Pi_{k}^{A}}\mid\psi_{A}\rangle\otimes|k\rangle)_{E}
\end{eqnarray}

The conditional mutual information $I(E;B\mid X)_{\tau_{XEB}}$ satisfy:
\begin{eqnarray}
I(E;B\mid X)_{\tau_{XBE}}&=&S(EX)_{\tau_{XEB}}+S(BX)_{\tau_{XEB}} \nonumber\\
&-&S(X)_{\tau_{XEB}}-S(EBX)_{\tau_{XEB}}
\end{eqnarray}

As an extension of quantum discord, the R\`{e}nyi quantum discord of $\rho_{AB}$ is defined for $\alpha\in(0,1)\cup(1,2]$ as [25]
\begin{eqnarray}
D_{\alpha}(\rho_{AB})=\mathop{inf}_{\Pi_{k}^{A}}I_{\alpha}(E;B\mid X)_{\tau_{XEB}}
\end{eqnarray}
where the  R\`{e}nyi conditional mutual information $I_{\alpha}(E;B\mid X)_{\tau_{XEB}}$ satisfy:
\begin{eqnarray}
I_{\alpha}(E;B\mid X)_{\tau_{XBE}}&=&\frac{\alpha}{\alpha-1}\log Tr\{(\rho_{X}^{\frac{\alpha-1}{2}}Tr_{E}\{\rho_{EX}^{\frac{1-\alpha}{2}}\rho_{EBX}^{\alpha}\nonumber\\
&&\rho_{EX}^{\frac{1-\alpha}{2}}\}\rho_{X}^{\frac{\alpha-1}{2}})^{\frac{1}{\alpha}}\}
\end{eqnarray}

The properties of the R\`{e}nyi quantum discord are shown in Table 2 of Ref.[25].
\subsection{\label{sec:level2}The Hamiltonian of the open system}
We consider two independent dimer systems which are coupled to two correlated
Fermi-spin environments, respectively, as shown in Fig.1. The Hamiltonian of the total system has the following form [27]:
\begin{eqnarray}
H=H_{d}+\sum_{i=1,2}H_{B_{i}}+\sum_{i,j=1,2}H_{d_{i}B_{j}}+q S_{1}^{z}S_{2}^{z}
\end{eqnarray}
where $H_{d}=H_{d_{1}}+H_{d_{2}}$ and $H_{B_{i}}$ describe two independent dimer system and Fermi-spin environments,
respectively. $H_{d_{i}B_{j}}$ and $S_{1}^{z}S_{2}^{z}$ denote the interaction between dimer system and that between environments
and the environments, respectively. The various parts of the Hamiltonian can be written as following forms:
\begin{eqnarray}
H_{d_{1}}&=&\varepsilon_{1}|1\rangle\langle1|+\varepsilon_{2}|2\rangle\langle2|+J_{1}(|1\rangle\langle2|+|2\rangle\langle1|)\nonumber\\
H_{d_{2}}&=&\varepsilon_{3}|3\rangle\langle3|+\varepsilon_{4}|4\rangle\langle4|+J_{2}(|3\rangle\langle4|+|4\rangle\langle3|)\nonumber\\
H_{d_{1}B_{1}}&=&\gamma_{1}|1\rangle\langle1|S_{1}^{z};H_{d_{1}B_{2}}=\gamma_{2}|2\rangle\langle2|S_{2}^{z}\nonumber\\
H_{d_{2}B_{1}}&=&\gamma_{3}|3\rangle\langle3|S_{1}^{z};H_{d_{2}B_{2}}=\gamma_{4}|4\rangle\langle4|S_{2}^{z}\nonumber\\
H_{B_{i}}&=&\alpha_{i}S_{i}^{z}\nonumber\\
\nonumber
\end{eqnarray}

\begin{figure}
\centering
\scalebox{0.3}[0.3]{\includegraphics{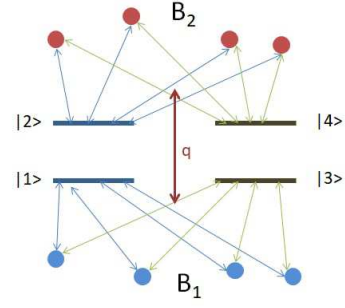}}
\caption{\label{fig:epsart} (Color online) The two dimer systems interacting with two interaction spin-environments ($S_{1}$ and $S_{2}$). The red and blue balls denote the spin particles of environments ($B_{1},B_{2}$), respectively. The $|1\rangle$ and $|2\rangle$ ($|3\rangle$ and $|4\rangle$) denote the energy level states of dimer system $H_{d_{1}}$ ($H_{d_{2}}$ ). $q$ denotes the interaction strength between the spin-environments $S_{1}$ and $S_{2}$. The arrows between the energy level states and spin particles of environments denote the interaction between dimer systems and environments.}
\end{figure}

Here, each environment $B_{i}$ consists of $N_{i}$ particles $(i=1,2)$ with spin $\frac{1}{2}$; $\varepsilon_{a}$ and $|a\rangle$ ($a=1,2,3,4$) are the energy levels and the energy states of the dimer system, $J_{1}$ and $J_{2}$ are the amplitudes of transition. The collective spin operators are defined as $S_{i}^{z}=\sum_{k=1}^{N_{i}}\frac{\sigma_{z}^{k,i}}{2}$, where $\sigma_{z}^{k,i}$ are the Pauli matrices and $\alpha_{i}$ is the frequency of $\sigma_{z}^{k,i}$. So $q S_{1}^{z}S_{2}^{z}$ describes an Ising-type correlation between the environments with strength $q$. The cases $q=0$ and $q\neq 0$, describe independent and correlated spin bath, respectively.
\subsection{\label{sec:level3}The dynamics evolution of the dimer system}
\quad The formal solution of the von Neumann equation ($\hbar=1$)
\begin{eqnarray}
\frac{d}{dt}\rho(t)=\mathcal{L}\rho(t)=-i[H,\rho(t)]
\end{eqnarray}
can be solved as
\begin{eqnarray}
\rho(t)=e^{\mathcal{L}t}\rho(0)
\end{eqnarray}
where $\rho(t)$ denotes the density matrix of the total system.

The dynamics of the reduced density matrix $\rho_{d}(t)$ is obtained by the partial trace method which discards the freedom of the environments. That is
\begin{eqnarray}
\rho_{d}(t)=Tr_{B}( e^{\mathcal{L}t}\rho(0))
\end{eqnarray}

Here, the states $|j,m\rangle$ denote the orthogonal bases in the environment Hilbert space $H_{B}$ which satisfy [28]:
\begin{eqnarray}
S^{2}|j,m\rangle&=&j(j+1)|j,m\rangle;\nonumber\\
S^{z}|j,m\rangle&=&m|j,m\rangle;S^{2}=(S^{x})^{2}+(S^{y})^{2}+(S^{z})^{2}\nonumber\\
j&=&0,...,\frac{N}{2}; m=j,...,-j\nonumber\\
\nonumber
\end{eqnarray}

For the initial state $\rho(0)=\rho_{d}(0)\otimes\rho_{B}(0)$ condition, the reduced density matrices $\rho_{d}(t)$ of the dimer system is
\begin{eqnarray}
\rho_{d}(t)&=&\frac{1}{Z}\sum_{j_{1}=0}^{N_{1}/2}\sum_{m_{1}=-j_{1}}^{j_{1}}\sum_{j_{2}=0}^{N_{2}/2}\sum_{m_{2}=-j_{2}}^{j_{2}}
\frac{\nu(N_{1},j_{1})\nu(N_{2},j_{2})}{e^{\beta q m_{1}m_{2}}e^{\beta\alpha_{1}m_{1}}e^{\beta\alpha_{2}m_{2}}}\nonumber\\
&&\times A^{\dagger}U^{\dagger}\rho^{'}_{d}(0)U A
\end{eqnarray}
where $\nu(N_{i},j_{i})$ denotes the degeneracy of the spin bath[28-29]. $\rho^{'}_{d}(0)$ is the matrix form of the density operator $\rho_{d}(0)$ under
the basis states of dimer system Hilbert space $A^{\dagger}=(\langle3|\langle1| \langle3|\langle2| \langle4|\langle1| \langle4|\langle2|)$. The
symbol $U$ in Eq.(12) denotes the $4\times4$ matrix and equal to $MBQ$ (here, $M$, $B$ and $Q$ are also $4\times4$ matrices [27]).

 In order to obtain Eq.(12), the environment is given as the canonical distribution
\begin{equation}
\rho_{B}(0)=\frac{1}{Z}e^{\beta q S_{1}^{z}S_{2}^{z}}\prod_{i=1}^{2}e^{-\beta\alpha_{i}S_{i}^{z}}\nonumber\\
\nonumber
\end{equation}
with $\beta=\frac{1}{K_{B}T}$($T$ is temperature and $K_{B}$ is Boltzmann constant). The partition function $Z$ is
\begin{eqnarray}
Z&=&\sum_{j_{1}=0}^{N_{1}/2}\sum_{m_{1}=-j_{1}}^{j_{1}}\sum_{j_{2}=0}^{N_{2}/2}\sum_{m_{2}=-j_{2}}^{j_{2}}
\frac{\nu(N_{1},j_{1})\nu(N_{2},j_{2})}{e^{\beta q m_{1}m_{2}}e^{\beta\alpha_{1}m_{1}}e^{\beta\alpha_{2}m_{2}}}\nonumber\\
\nonumber
\end{eqnarray}
\subsection{\label{sec:level4}The properties of R\`{e}nyi discord}
\quad In this section, the changing behaviors of the quantum correlation are discussed for the two-qubit X [9] and special canonical initial (SCI)[23] under different parameters, respectively. The two-qubit X state is widely used in condensed matter systems and quantum dynamics[9-10,19,30-31]. Under the basis vectors $|00\rangle$, $|01\rangle$, $|10\rangle$ and $|11\rangle$ ( here $0$($1$) denotes the spin up (down) state ), the density matrix of a two-qubit X state can be written as
\begin{eqnarray}
\rho_{d}(0)=\left[
\begin{array}{cccc}
 a     & 0 &0 &\delta\\
 0    & b &\beta &0\\
 0    & \beta^{*}  &c &0\\
 \delta^{*}      & 0  &0  &d\\
\end{array}
\right]
\end{eqnarray}
satisfying $a,b,c,d\geq0$,$ a+b+c+d=1$, $||\delta||^{2}\leq ad$ and $||\beta||^{2}\leq bc$.

\quad Unlike X states, the class of canonical initial (CI)states [23] have the density matrix
\begin{eqnarray}
\rho_{d}(0)=\frac{1}{4}\left[
\begin{array}{cccc}
 1+C_{33}     & C_{01} &C_{10} &C_{11}-C_{22} \\
 C_{01}^{*}    & 1-C_{33} &C_{11}+C_{22} &C_{10}\\
 C_{10}^{*}   & C_{11}+C_{22}  &1-C_{33} &C_{01}\\
 C_{11}-C_{22}      & C_{10}^{*}  &C_{01}^{*}  &1+C_{33}\\
\end{array}
\right]\nonumber\\
\end{eqnarray}
and the SCI states satisfy:
\begin{eqnarray}
\left\{
\begin{array}{rcl}
&& C_{22}/C_{33}=-C_{11}\\
&& C_{10}/C_{01}=C_{11}\\
&&(C_{33})^{2}+(C_{01})^{2}\leq 1\\
\end{array} \right.
\end{eqnarray}

In view of the freezing phenomenon for X (SCI) states [9-10,14,19,23,30-31] by geometry and von Neumann entropy discords, X and SCI initial states are chosen here.

\begin{figure}

\centering

\subfigure{\label{fig:subfig:a}

\includegraphics[width=1\linewidth]{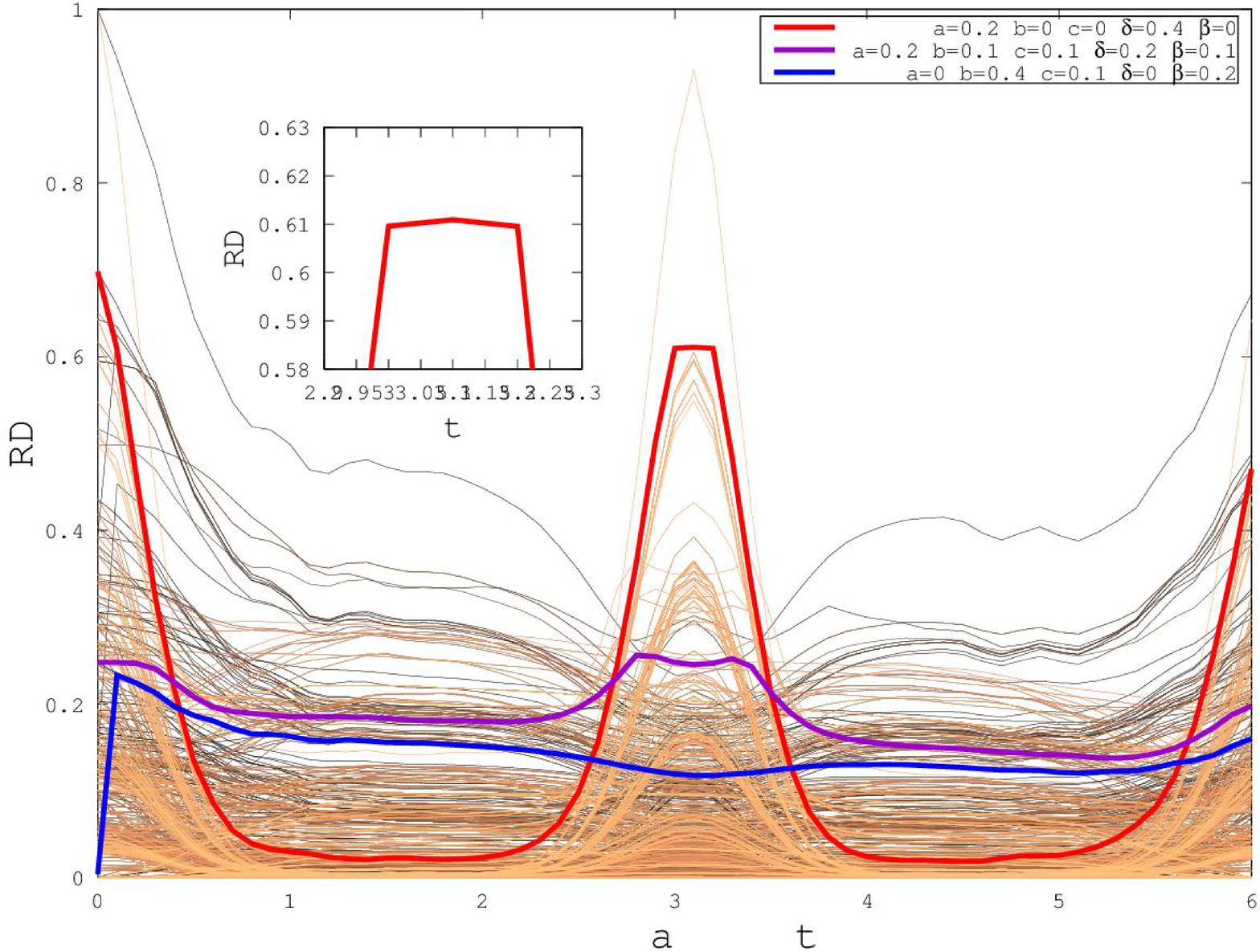}}

\hspace{0.01\linewidth}

\subfigure{\label{fig:subfig:a}

\includegraphics[width=1\linewidth]{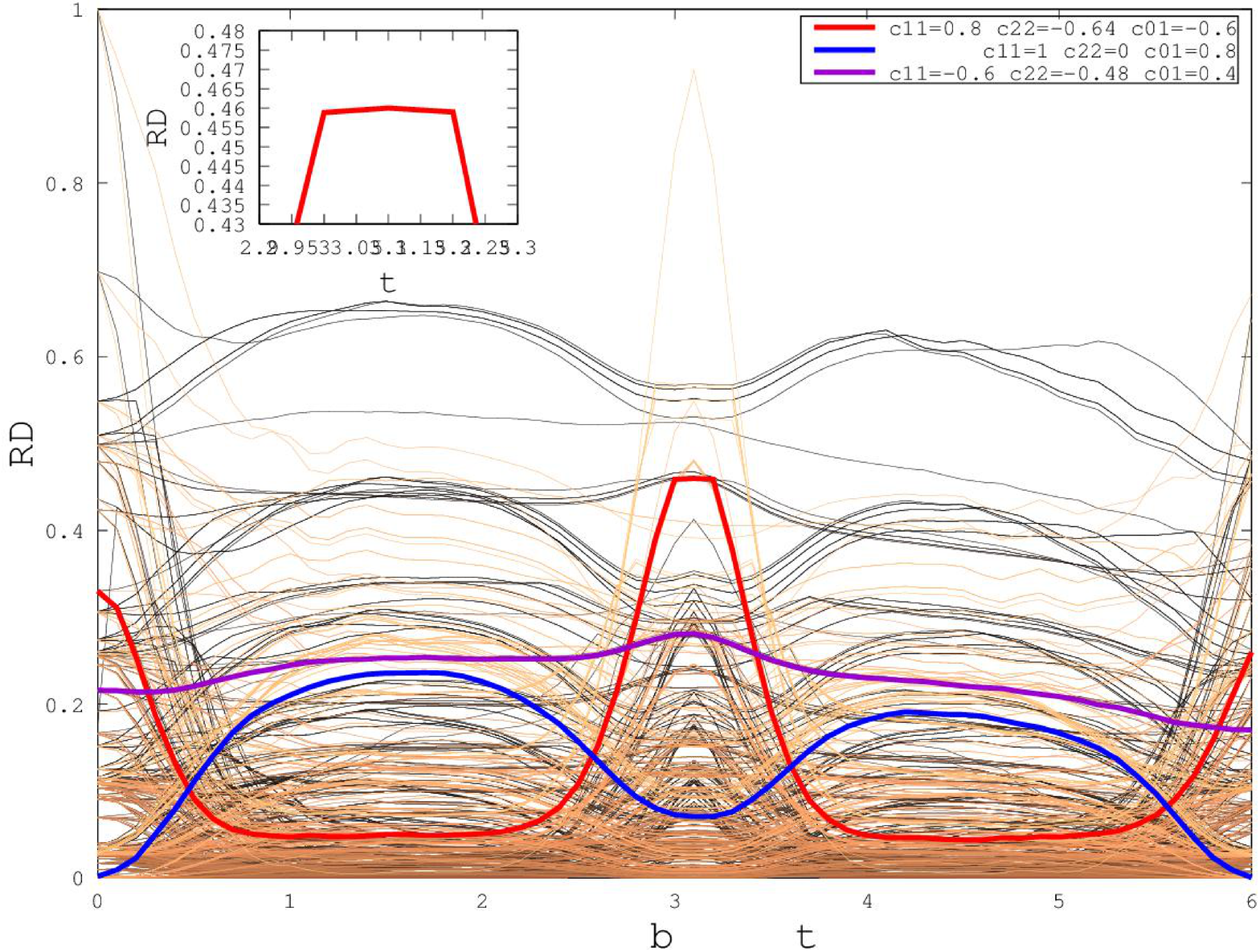}}

\hspace{0.01\linewidth}
\caption{\label{fig:epsart} (Color online) The properties of R\`{e}nyi discord as a function of time $t$ for X and SCI initial states. The parameters are $\alpha_{1}=250 ps^{-1}$, $\alpha_{2}=200 ps^{-1}$, $\Delta_{1}=20 ps^{-1}$, $\Delta_{2}=10 ps^{-1}$,$\Delta_{3}=22 ps^{-1}$,$\Delta_{4}=12 ps^{-1}$, $q=30ps^{-1}$, $\beta=1/77$, $N_{1}=20$, $N_{2}=22$ , $\gamma_{1}=1 ps^{-1}$,$\gamma_{2}=1.1 ps^{-1}$,$\gamma_{3}=0.9 ps^{-1}$,$\gamma_{4}=1.2 ps^{-1}$, $J_{1}=10 ps^{-1}$,$J_{2}=12 ps^{-1}$ and $\alpha=0.9$.}
\end{figure}

In Fig.2, the changing behaviors of quantum correlation are shown for X (Fig.2(a)) and SCI(Fig.2(b)) initial states. With the time evolution, the quantum correlation displays the nom-Markov behaviors, especially the peak at about 3 second for some initial states. It indicates the feedback of quantum information (quantum correlation) with the nom-Markov process. There are the freezing phenomena for some initial states, with the purple and red solid lines denoting the freezing phenomena for two initial states, respectively. It also hints that the freezing phenomenon of quantum correlation is a universal quantum character and has a deep physical meaning. In the inset box, it shows the partially enlarged drawing of the max freezing quantum correlation. Simultaneously, the blue solid line shows the spring of the quantum correlation for the initial states with zero quantum correlation at $t=0$. It means that we can generate the quantum correlation by the environments of this quantum system.

\begin{figure}

\centering

\subfigure{\label{fig:subfig:a}

\includegraphics[width=1\linewidth]{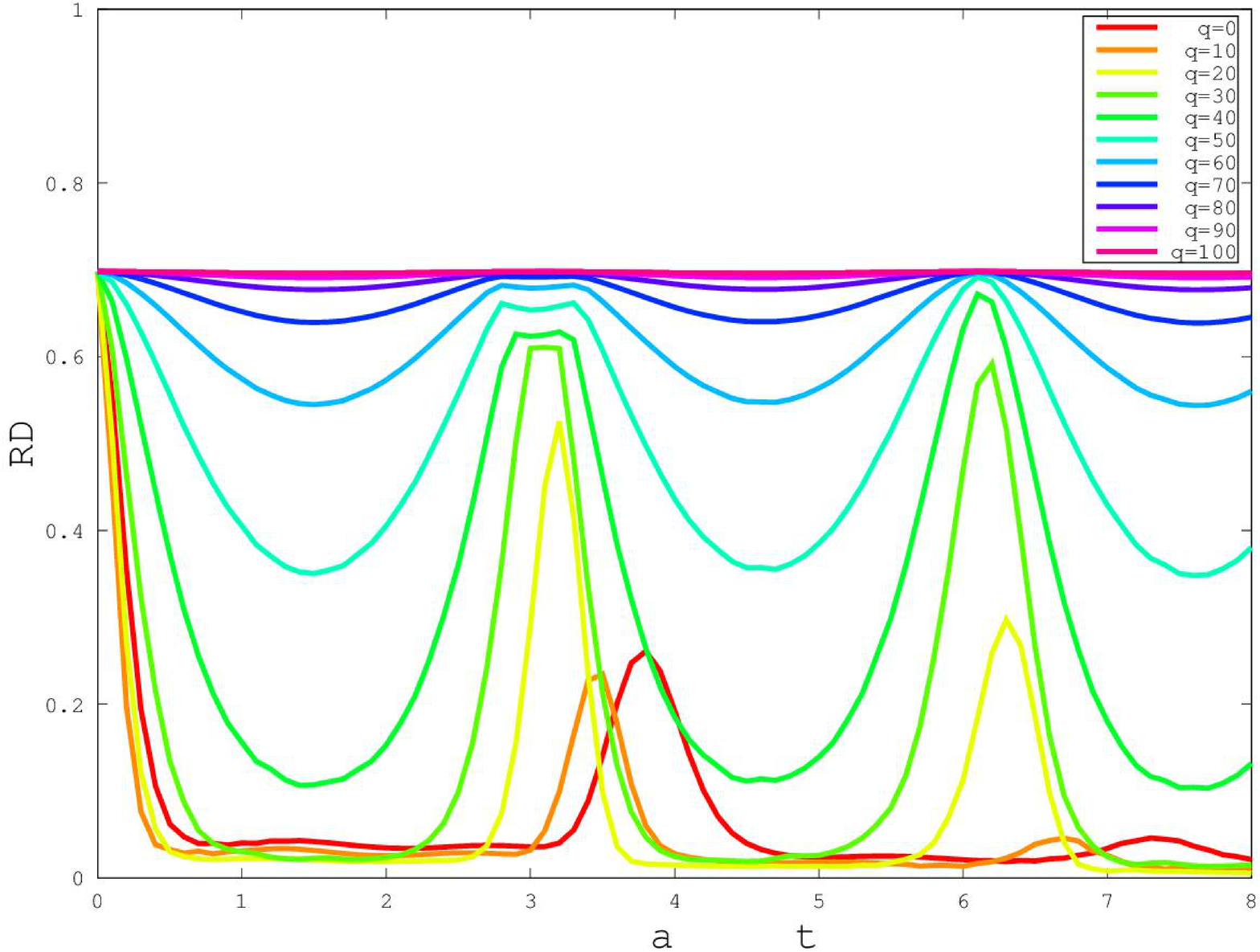}}

\hspace{0.01\linewidth}

\subfigure{\label{fig:subfig:a}

\includegraphics[width=1\linewidth]{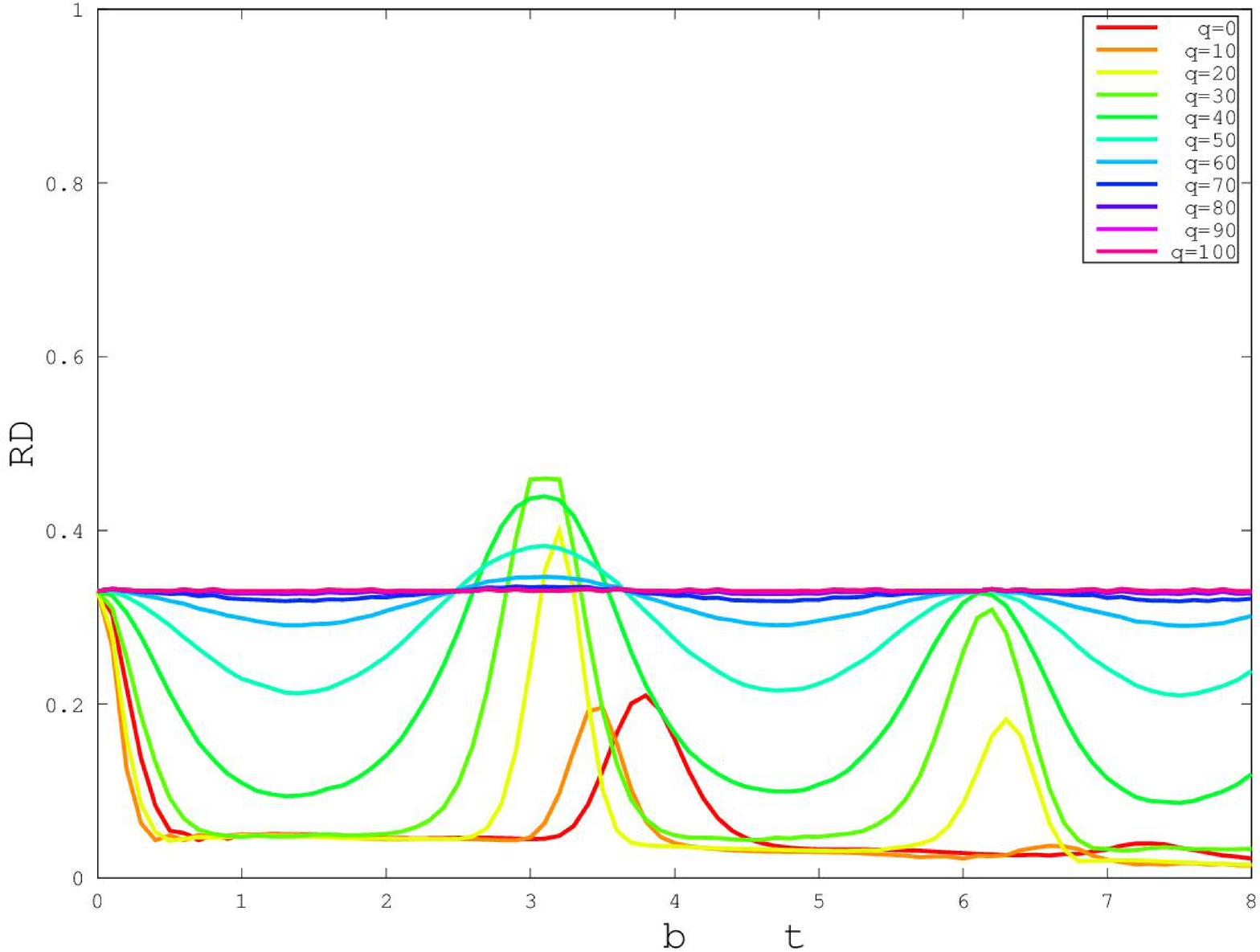}}

\hspace{0.01\linewidth}
\caption{\label{fig:epsart} (Color online) The changes of freezing R\`{e}nyi discord based on X (Fig.3(a)) and SCI (Fig.3(b)) initial states (blue solid line of Fig.(1)) with time $t$ for different environment coupling parameters $q$. The other parameters are same in Figure 2.}
\end{figure}

Although the general results are obtained, the intrinsic parameters may play an important role in the changing behaviors of the quantum correlation [14-15,17,32], especially the freezing behavior. In Fig.3, the environment coupling parameter $q$ can strongly affect the occurrence of freezing phenomenon, and the quantum correlation appears the quasi-periodic oscillations for X (Fig.3(a)) and SCI(Fig.3(b)) initial states. However, the oscillation behaviors are depressed with $q\geq 50 $ for X and SCI initial states. The freezing phenomena of quantum correlation also spring up with increasing $q$ for X state. But, the freezing phenomena of SCI state first appears with increasing $q$, and then disappears with $q\geq 70$. Particularly, the freezing phenomena always exist throughout the time when $q$ excesses 90. Furthermore, the value of quantum correlation increases with increasing $q$ for X state. From the perspective of the non-Markovian dynamical process, the larger $q$ means the more information flowing from the system into the environment than that from the environment into the system. Therefore, a reasonable value of $q$ is important for the maintenance of quantum correlation.

\begin{figure}

\centering

\subfigure{\label{fig:subfig:a}

\includegraphics[width=1\linewidth]{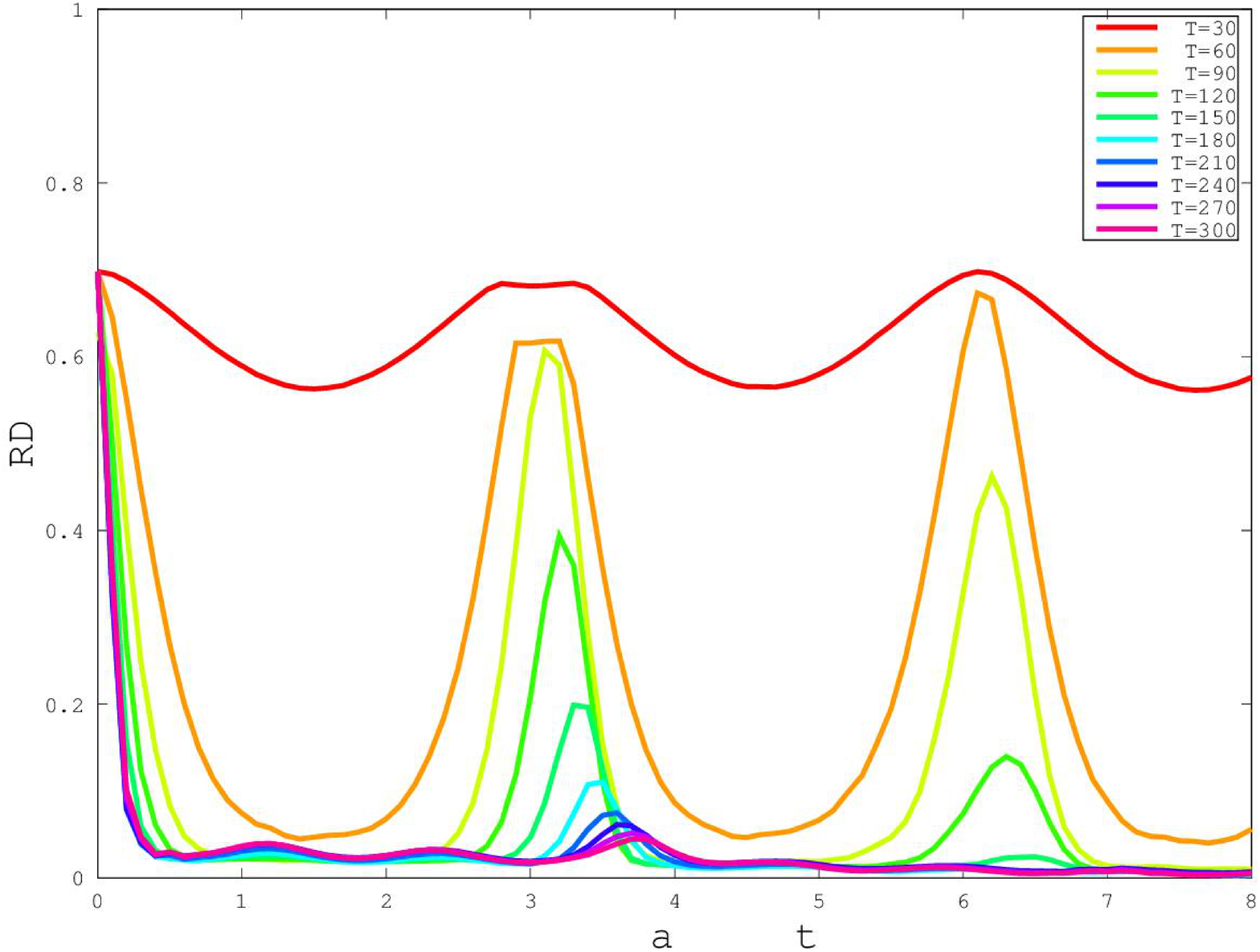}}

\hspace{0.01\linewidth}

\subfigure{\label{fig:subfig:a}

\includegraphics[width=1\linewidth]{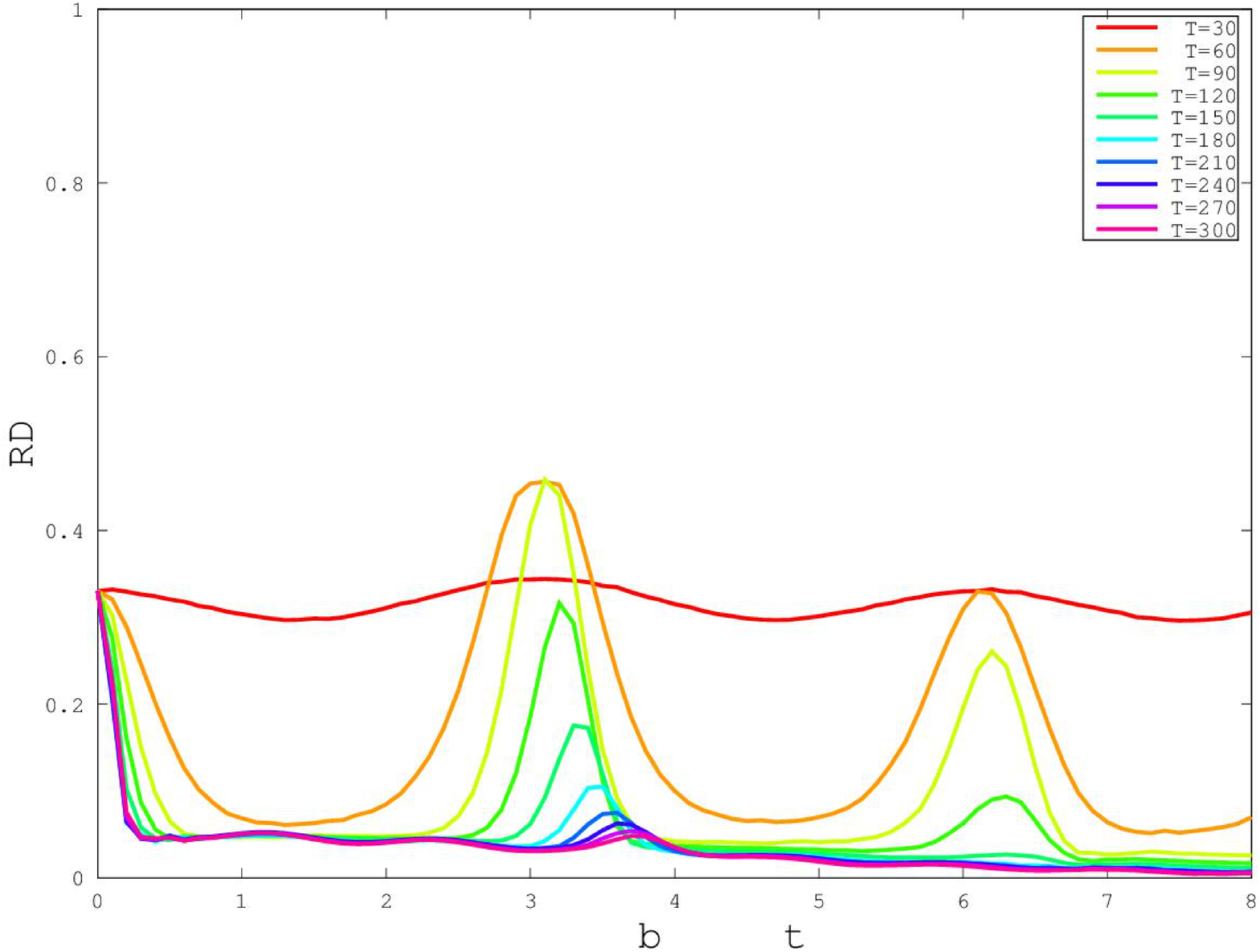}}

\hspace{0.01\linewidth}
\caption{\label{fig:epsart} (Color online) The changes of freezing R\`{e}nyi discord based on X (Fig.4(a)) and SCI (Fig.4(b)) initial states (blue solid line of Fig.(1)) with time $t$ for different environment temperatures $T$. The other parameters are same in Figure 2.}
\end{figure}

Except for the parameter $q$, the temperature T is also important for the quantum correlation. According to our previous works [24,25], a higher temperature may depress the activity of quantum correlation. How does temperature affect the frozen platform? The effect of temperature on the frozen platform is shown in Fig.4. With increasing temperature $T$, the frozen platform collapses and then reappears at $T\geq 150$. Simultaneously, the platform height is reduced.
\begin{figure}

\centering

\subfigure{\label{fig:subfig:a}

\includegraphics[width=1\linewidth]{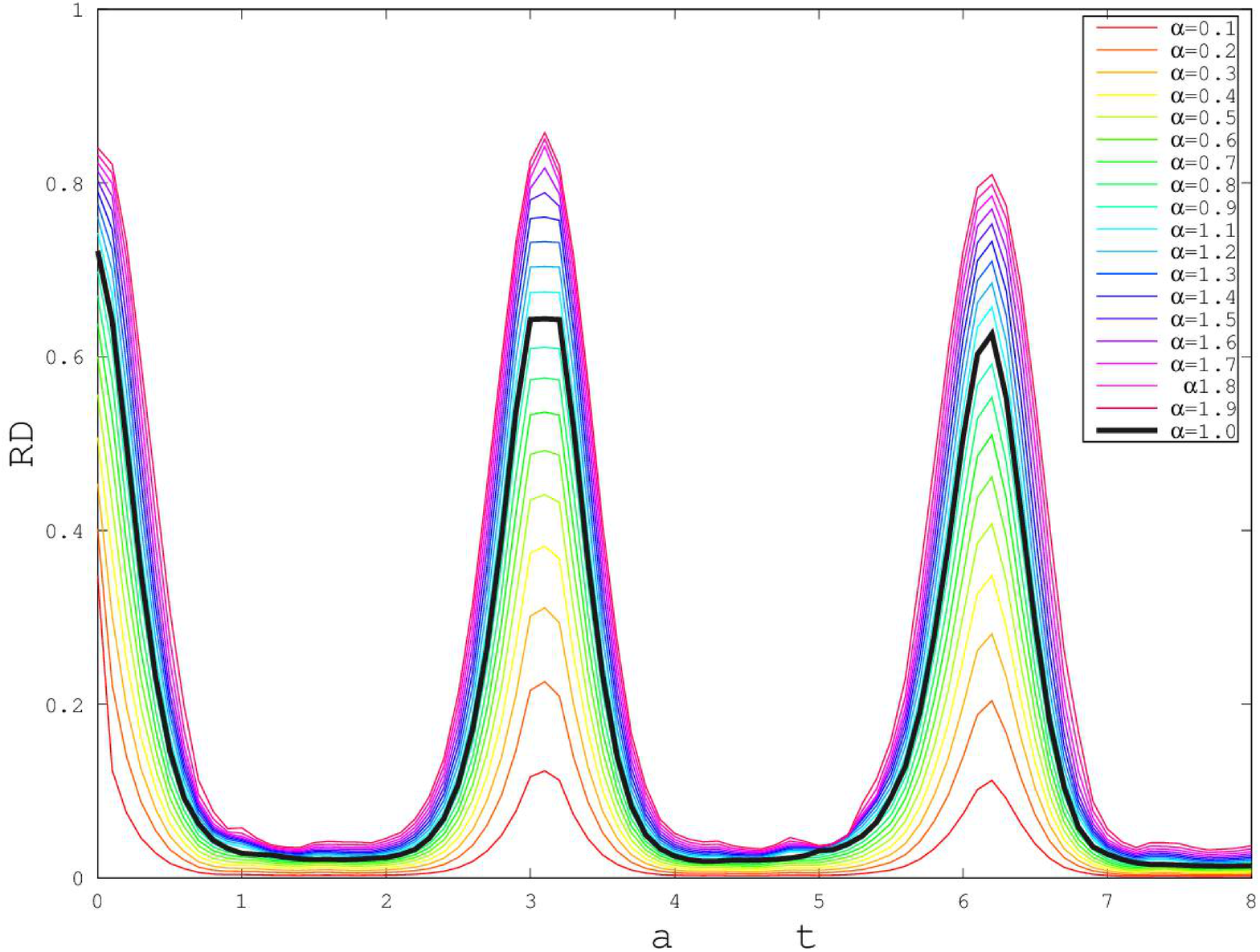}}

\hspace{0.01\linewidth}

\subfigure{\label{fig:subfig:a}

\includegraphics[width=1\linewidth]{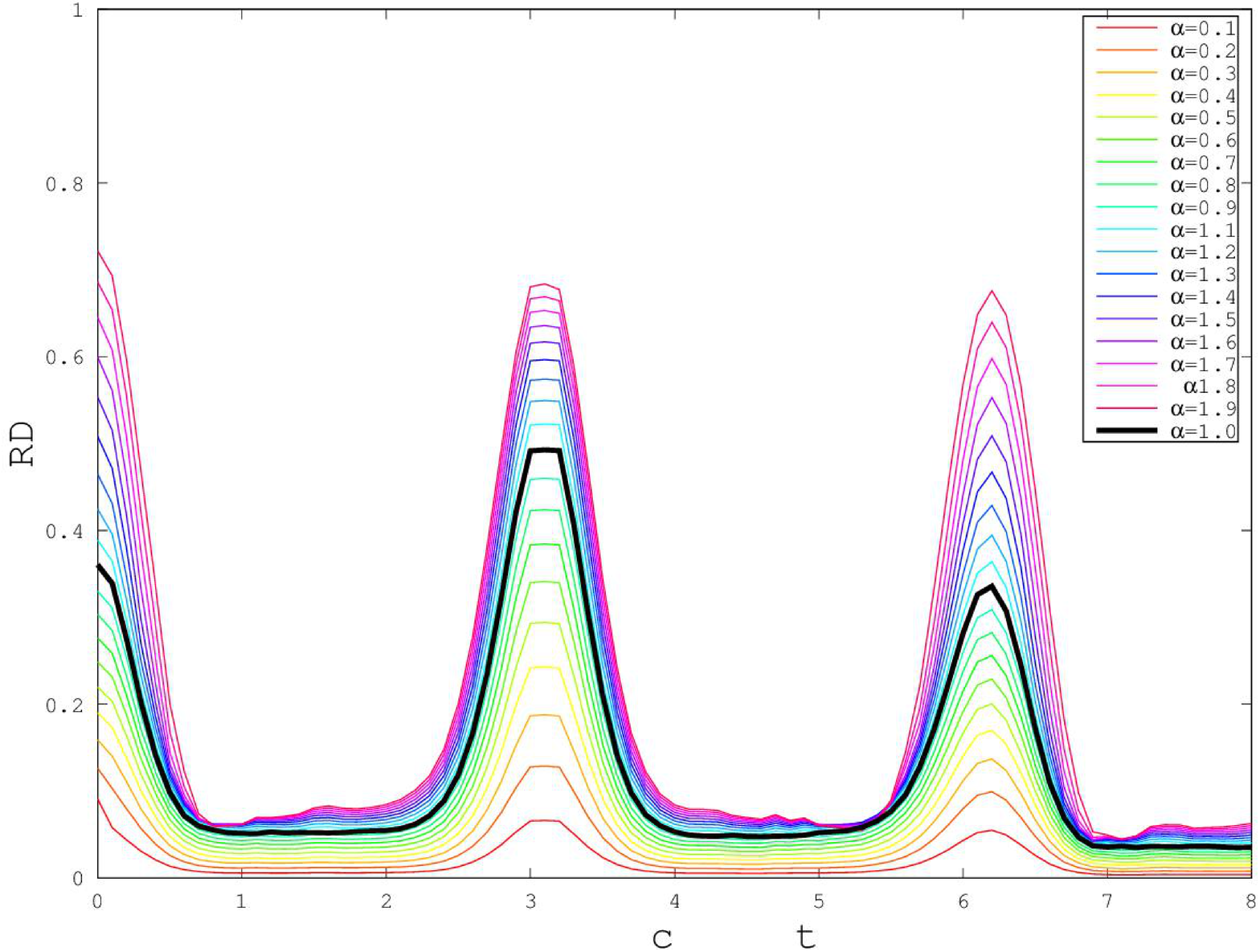}}

\hspace{0.01\linewidth}
\caption{\label{fig:epsart} (Color online) The changes of freezing R\`{e}nyi discord based on X (Fig.4(a)) and SCI (Fig.4(b)) initial states (blue solid line of Fig.(1)) with time $t$ for different parameter $\alpha$. The other parameters are the same as Figure 2.}
\end{figure}

Fig. 5 display the effect of different parameters $\alpha$ which is an important parameter of R\`{e}nyi entropy for R\`{e}nyi discord. With the increase of $\alpha$, the monotonicity of R\`{e}nyi discord is well displayed [25]. For X and SCI initial states, the freezing platform appears in the range of $\alpha \in [0.7,1.4]$ and $\alpha \in [0.1,1.6]$, respectively.  When the freezing platform appears, there is a particular scope of parameter alpha, which depends on the initial states. Finally, compared the black line ($\alpha =1$) with others, the quantum discord only shows part of the nature of quantum correlation which quantifies by one of entropy discord, while the others correspond to different entropy discord.  This otherness maybe supply the help to discuss the difference between quantum discord and geometric discord, especially, the occurrence of freezing phenomenon has different conditions.

\section{\label{sec:level1}Conclusion}
In this paper, the changing properties of R\`{e}nyi discord are shown for two independent Dimer System  coupled to two correlated Fermi-spin environments.  Three main results are presented: 1) the freezing behaviors still exist for R\`{e}nyi discord which is not just a mathematical coincidence. 2) the freezing platform depends on the parameter $\alpha$ under the same conditions due to the divergence and nonlinearity properties of R\`{e}nyi entropy. 3) for larger parameter $q$ and lower temperature $T$, the collapse of the freezing quantum correlation is depressed. These results would supply help to the measurement of quantum correlation and the research of quantum information.

\end{document}